# GUI CONTROL SYSTEM FOR THE MU2E ELECTROSTATIC SEPTUM HIGH VOLTAGE AT FERMILAB*

R. Kibbee†, E. Prebys, University of California Davis, Davis, CA, USA
V. Nagaslaev, Fermi National Accelerator Laboratory, Batavia, IL, USA

## Abstract

The Mu2e Experiment has stringent beam structure requirements; namely, its proton bunches with a time structure of 1.7 µs in the Fermilab Delivery Ring. This beam structure will be delivered using the Fermilab 8-GeV Booster, the 8-GeV Recycler Ring, and the Delivery Ring. The 1.7-µs period of the Delivery Ring will generate the required beam structure by means of a third order resonant extraction system operating on a single circulating bunch.

The electrostatic septum (ESS) for this system is particularly challenging, requiring mechanical precision in a ultra high vacuum of $1 \times 10^{-8}$ Torr to generate 100 kV across 15 mm. This paper describes a graphical user interface that has been developed to automate the conditioning and commissioning process for the electrostatic septa. It is based on an interface to the Fermilab ACNET system using the ACSys Python Data Pool Manager (DPM) Client produced and maintained by Fermilab Accelerator Controls.

Network interfacing between data pool managers made by the application and ACNET devices introduce an inherent (approximately 1 s) latency in throughput of the readouts. This delay is utilized to process and graph incoming data events of devices crucial to conditioning of a electrostatic septum (ESS). 'Ramping' and 'Monitoring' modes adjust settings of the power supply based on internal logic to efficaciously increase and maintain the high voltage (HV) in the ESS, easing the voltage setting on incidence of sparking or other possibly damaging events. A timestamped log file is produced as the application runs.

## INTRODUCTION

The requirement for conditioning is deeply related to the surface of the cathode. The internal configuration of the septum can be seen in Fig. 1. Two electrostatic septa will be used in the final configuration of the resonant extraction system in the Delivery Ring. The areal dimensions for the upstream (ESS1) cathode are 1.856 in by 52.598 in while the downstream (ESS2) cathode has dimensions 1.856 in by 65.500 in. This resulting in a total cathode surface area of 629.817 cm$^2$ and 784.308 cm$^2$ respectively for ESS1 and ESS2. While both cathodes have undergone extensive cleaning and polishing in a vacuum environment to minimize incidence of surface impurities both chemical and structural, the large areal surface of the cathode allows a substantial occurrence of imperfections. One concern is that electric field lines are always perpendicular to the surface of conductors,
mechanical imperfections such as micro-protrusions in the high voltage environment act as sites of high density for the electric field. At high enough voltages, micro-protrusions will have electric field emission of electrons within, this emission may lead to a large spark.

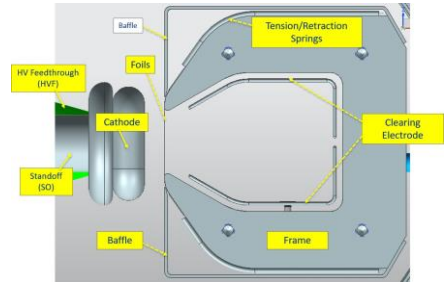

Figure 1: Internal configuration of the electrostatic septa.

These sparks and those from other causes can be destructive in beneficial and detrimental ways. Using the example of sparks caused by field emission on micro-protrusions, those that are beneficially destructive have only the micro-protrusion causing the spark being flattened without the surrounding area significantly affected. The detrimentally destructive sparks have more than just the micro-protrusion affected, producing more and possibly larger micro-protrusions in the area of the spark origin. The degree to which either incident occurs on these spark events and others is influenced by how the ramping of the voltage on the ESS HV power supply is executed. A sizeable, sudden increase of the voltage is more probable to cause a damaging spark, whereas slower, more incremental increases to the voltage will help to ensure the sparks work to better condition the ESS by removing imperfections. Thus conditioning the electrostatic septa requires a means of incrementally ramping the voltage on the HV power supply at a desirable pace.

The ESS HV power supply is connected to Fermilab's ACNET system, enabling authorized users to access the power supply from an ACNET console and adjust settings to change the voltage. There are also sensors as part of the resonant extraction system to enable direct readouts of the voltage from the HV power supply, the sum current from the ESS and its HV power supply, and vacuum readings within the ESS vessel. A spark counter is also implemented giving an integer count of sparking events based on sudden voltage drops on the HV cable of the ESS. These parameters enable a trained user to monitor the values from each sensor while ramping the voltage to condition the ESS, such that the ESS can achieve its desired voltage without concern of damaging sparks breaking down the vacuum or vitiating ESS components.

---

* This work has been supported by US Department of Energy Grant DE-SC0009999
† rbkibbee@ucdavis.edu

The process of incrementally ramping the voltage is time consuming and difficult, however, so it is preferable to automate the process with an application. This is possible using the recently developed ACSys Python data pool manager (DPM) client developed at Fermilab by the Fermi-Controls group. Using the package enables interfacing with Fermilab data acquisition and controls of devices on the ACNET system. This, together with the PyQt5 Python package for using the Qt graphical user interface framework and implementing established logic for ramping and monitoring the ESS parameters described above, enabled development of a GUI control system for the conditioning of the electrostatic septa.

## ACSYS DPM AND PYQT5

The ACSys DPM Client produces a DPM capable of interfacing with Fermilab's ACNET system and the devices therein [1]. Events with information including the device designation, timestamp, and readout values are taken from ACNET to the DPM and then extracted and processed as defined within the written code. This communication between ACNET, the DPM, and the application requires time and thus introduces a latency of about 1 s before the application obtains events with information of readouts from the ACNET devices. This is comparable to human response time and is accounted for in the timed calls to the DPM, enabling consistent throughput and time for the application to graph readouts as functions of time as well as run the readout values through the logic to determine a response. A list of ACNET devices and their measurement type is provided in Table 1.

Table 1: Measurements obtained from ACNET Devices.

| ACNET Device | Measurement Type |
| --- | --- |
| D:ESS2 | Voltage Reading |
| D_ESS2 | Voltage Setting |
| D:ESS2I | Current Reading |
| D_ESS2I | Current Protection Setting |
| D:IPES2A | Pressure Reading |
| D:IPES2B | Pressure Reading |
| D:IPES2C | Pressure Reading |
| D:IGES2 | Pressure Reading |
| D:ESS2SR | Spark Count |

PyQt5 is a set of Python bindings for the Qt application framework from The Qt Company [2], enabling Python applications to use the Qt graphical user interface (GUI) development capabilities. This includes plotting capabilities to graph the readout value and associated timestamp of each event as well as various widgets that can act as inputs, outputs, and data displays for a comprehensive application of the systems necessary for conditioning the electrostatic septa.

A full view of the GUI control system graphing device readouts is shown in Fig. 2.

## LOGIC

The GUI has a 'Ramping' mode for incrementally increasing the HV power supply voltage setting (D_ESS2) and a 'Monitoring' mode for holding the voltage at a set value given by the "HV Target" input. Should define events occur both modes will lower the voltage and 'Ramping' mode will pause or stop incrementation of the HV setting. These events are based on inputs visible in the lower right section of Fig. 2.

A pause causes the voltage to be set back by the value inputted in the "Ease" input and will stop the 'Ramping' mode for the amount of time in the "Pause Time" input. In 'Monitoring' mode, the voltage is eased and then paused for the same time until ramping back to the value set for further monitoring. Pauses occur on three conditionals: the spark count increases from its prior value; the difference in readout values between the voltage setting and reading devices exceeds the "Sag Thresh" input, also known as the voltage sag threshold; or the vacuum readout value designated for the processing exceeds the value inputted in "Vac Thresh".

A stop for 'Ramping' mode eases the voltage as with the pause but then turns off the 'Ramping' mode and sets the application into 'Monitoring' mode. A stop for 'Monitoring' mode reduces the "HV Target" input about which the HV power supply setting is held. Stops occur on four conditionals: the spark count exceeds the spark threshold set by the initial spark count added to the "Spark Thresh" input; the voltage reading exceeds the "HV Target" input, as it is expected that the reading value will be lower than the setting value; the time value exceeds the "Stop Time" input; and if there is excessive pausing.

As incidents necessitating stops are more pressing than pauses, the logic puts occurrences necessitating stops as higher priority than pauses.

A log file is produced including the initial start time of the application, inputs used in the logic of the application at the initial state and including updates, and messages providing details during active use of the GUI application in 'Monitor' and 'Ramping' modes. These messages include causes for stops and pauses, time stamps for when settings will be made, and notifications if they are made.

## CONCLUSION

A graphic user interface control system for the high voltage power supply and other supporting devices in the electrostatic septa of the resonant extraction system has been developed and discussed in this paper. This tool will be of use for finalizing the commissioning of the prototype ESS installed in Fermilab's Delivery Ring and commissioning the second ESS to be added for the final configuration of the resonant extraction system.

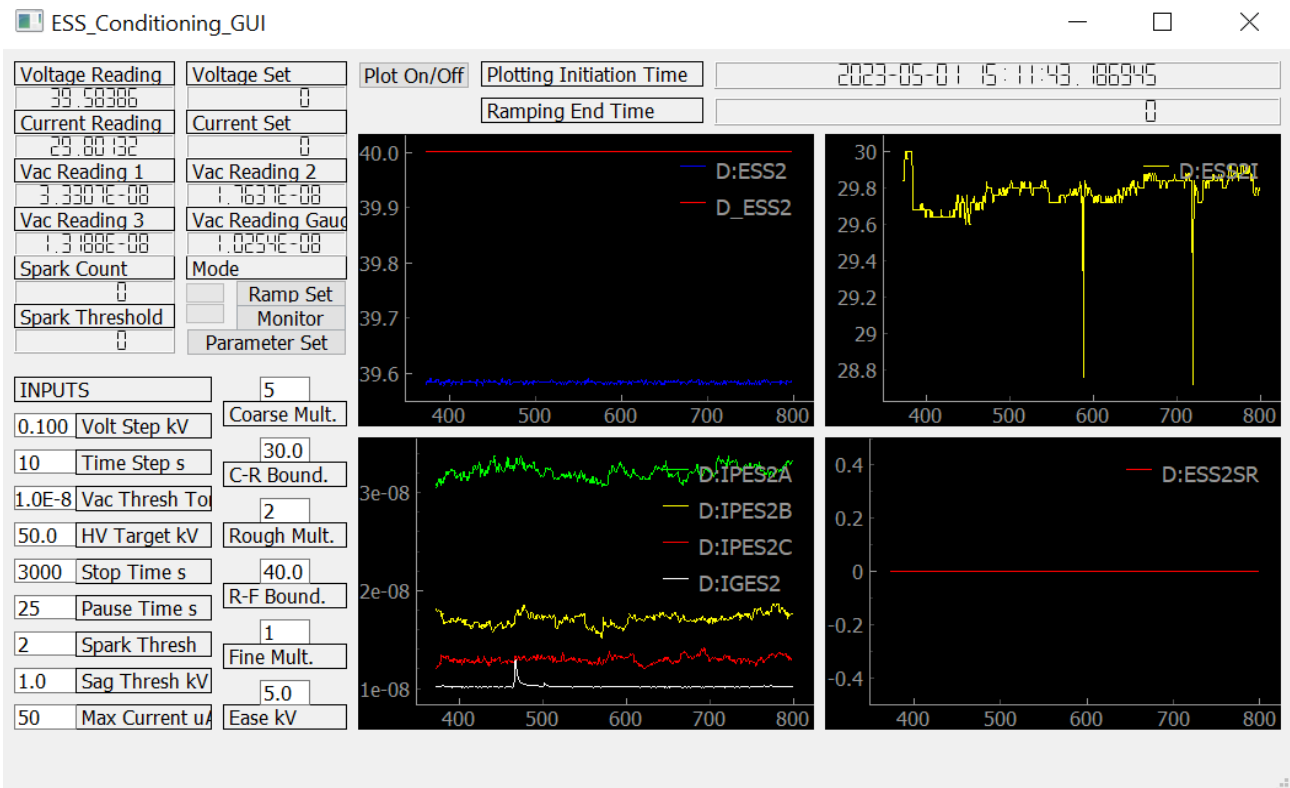

Figure 2: Screenshot of GUI control system graphing readouts of the ACNET devices critical to the conditioning of ESS2.